\documentclass[english,10pt]{article}
\usepackage[T1]{fontenc}
\usepackage[latin9]{inputenc}
\usepackage[a4paper]{geometry}
\usepackage{babel}
\usepackage{float}
\usepackage{url}
\usepackage{amsmath}
\usepackage{graphicx}
\usepackage{setspace}
\usepackage[unicode=true]{hyperref}

\makeatletter

\providecommand{\tabularnewline}{\\}

\usepackage{cite}

\makeatother

\title{Illusion of persistence in NBA 1995-2018 regular season data}
\author{A. Kononovicius}
\date{}

\begin{document}

\maketitle

\begin{abstract}
Among the sports fans beliefs about ``hot hands'' and ``winning streaks''
are widely spread, while the scientific debate about these effects
is still ongoing. Recently in a paper by P.Ferreira {[}\href{https://doi.org/10.1016/j.physa.2018.02.050}{Physica A 500: 92--96}{]}
detrended fluctuation analysis was applied to the NBA teams' win records.
It was shown that 28 considered NBA teams exhibit persistence in the
win record time series. In this paper we take the same data set and
compare the obtained results against various random models. We find
that the empirical results are consistent with the results obtained
from various simple random models.
\end{abstract}

\section{Introduction}

Numerous sports fans as well as majority of professional players tend
to believe in ``hot hand'' and ``winning streak'' phenomena. These
terms refer to the belief that performance of a player or a team might
become significantly better in comparison to their average\cite{Gilovich1984,Gilovich1985CogPsy,BarEli2006}.
During the last 3 decades there were significant number of attempts
to find strong statistical evidence for these beliefs as well as to
explain why do people believe in these phenomena \cite{Ayton2004MemCog,BarEli2006,Gilden1995CogPsy}.
Unsurprisingly most of the statistical approaches have failed to find
any evidence for these beliefs and those that found evidence were
shown to be inconclusive \cite{Vergin2000,BarEli2006,Green2016,Ferreira2018PhysA}.
While the cognitive psychologists tend to explain these beliefs by
the inability to recognize that small samples are not a good representation
of the population as whole \cite{Gilden1995CogPsy,Ayton2004MemCog}.
Though there are a few recent papers with criticism towards some of
the most well established analyses in the field, e.g., in \cite{Miller2016SSRN}
it is argued that the analysis carried out in \cite{Gilovich1985CogPsy}
suffers from selection bias and that after correcting for this bias
the conclusions of the original analysis reverses: the evidence confirming
``hot hand'' phenomenon is found. There is still no consensus in the
field and the debate on the apparently simple topic is still going
on. Though at this time the majority of literature on the topic is
dedicated to the individual performance in team sports (mostly basketball
and baseball), the papers on the teams performance are significantly
rarer.

In the last few decades the study of complex systems has developed
variety of tools to detect various anomalous features inherent to
complex systems. One of the thoroughly studied features of the complex
systems is persistence, which is assumed to indicate the presence
of memory in the time series. The ``hot hand'' and ``winning streak''
phenomena are exactly about the temporal persistence of good (or bad)
results of a player or a team. Hence it is tempting to check whether
sports time series exhibit such persistence. Detecting persistence
could potentially serve as a proof for the existence of these phenomena
as well as have other implications outside the scientific study. As
in \cite{Ferreira2018PhysA}, we will also rely on methodology known
as detrended fluctuations analysis (DFA), which is used to detect
persistence in various time series. Using DFA technique it was shown
that various economic, social and natural systems exhibit some degree
of persistence \cite{Ivanova1999PhysA,Kantelhardt2001PhysA,Ausloos2001PhysRevE,Kantelhardt2002PhysA,Ausloos2002CompPhysComm,Peng2004PhysRevE,Kwapien2005PhysA,Lim2007PhysA,Shao2012,Rotundo2015PhysA,Chiang2016PlosOne}.
Showing that sports time series also exhibit persistence could lead
to a better forecasting in sports as well as better understanding
of the persistence phenomenon in general. To supplement DFA results
we also employ first passage time methodology as well as study autocorrelation
functions (ACFs) of the series. The data itself invites the use of
first passage time methodology as the lengths of the streaks are by
definition identical to the first passage times, while ACFs is another
way to explore persistence in the data series.

We have organized the paper as follows: in Section~\ref{sec:analysis-original}
we analyze the original data, in Section~\ref{sec:analysis-shuffle}
we extend the previous analysis by considering the shuffled data as
well as data generated by the random models, while conclusions are
given in Section~\ref{sec:Conclusion}.

\section{Analysis of the original data\label{sec:analysis-original}}

As in \cite{Ferreira2018PhysA} we have downloaded NBA team regular
season (from 1995 to 2018) win records for $28$ NBA teams from \href{http://www.landofbasketball.com}{landofbasketball.com}
website. We have excluded NOP (New Orleans Pelicans) and CHA (Charlotte
Hornets/Bobcats) from the analysis, because these teams did not participate
in some of the considered seasons. Yet the games NOP and CHA played
against other teams are present in the other team records. Hence all
$28$ teams under consideration have played $1838$ games. We have
coded their victories as $+1$, losses as $-1$, while a single game
between BOS (Boston Celtics) and IND (Indiana Pacers), which was cancelled
due to Boston marathon bombings, was coded as $0$.

\begin{figure}
\begin{centering}
\includegraphics[width=0.4\linewidth]{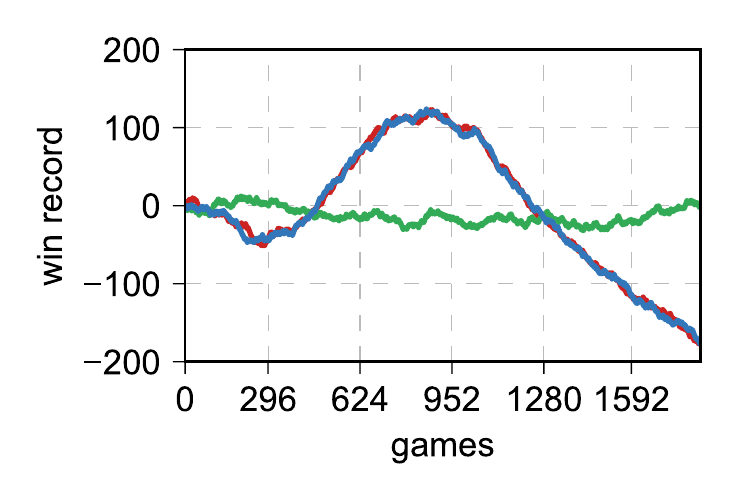}
\par\end{centering}
\caption{(color online) Win record for SAC (red curve), single example of the
fully shuffled win record for SAC (green curve) and single example
of the in-season shuffled win record for SAC (blue curve).\label{fig:record-example}}

\end{figure}

As an example of how the empirical series look like in Fig.~\ref{fig:record-example}
we have plotted win record series for SAC (Sacramento Kings). Additional
curves in the figure show the examples of shuffled series. We will
refer back to this figure when discussing shuffling algorithms in
a more detail in the later sections of this paper.

After running the original data through DFA, we have obtained results
similar to those reported in \cite{Ferreira2018PhysA}. We have found
that $26$ of the $28$ teams exhibit persistent behavior as their
$H>1/2$. Only DAL (Dallas Mavericks) and OKC (Oklahoma City Thunder/Seattle
Supersonics) are within margin of error from $H=1/2$. For the detailed
results see Table~\ref{tab:dfa-original}.

\begin{figure}
\begin{centering}
\includegraphics[width=0.7\textwidth]{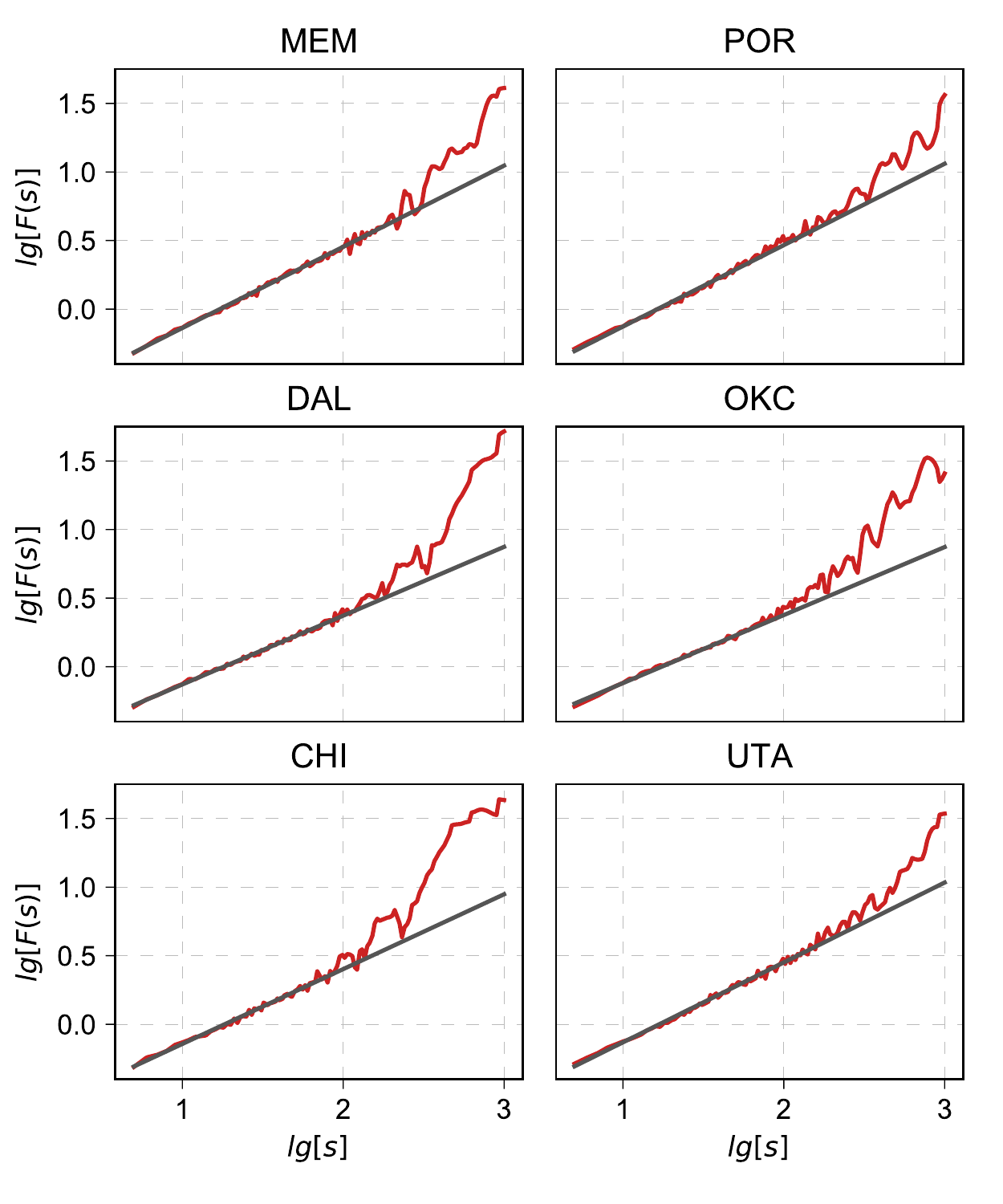}
\par\end{centering}
\caption{(color online) Quality of the fits, $F(s)\sim s^{H}$, provided by
the reported Hurst exponents, Table~\ref{tab:dfa-original}, in the
six selected cases: MEM and POR have the highest $H$ estimates, DAL
and OKC have the lowest $H$ estimates, while CHI and UTA were used
as examples in \cite{Ferreira2018PhysA}.\label{fig:original-extreme}}
\end{figure}

\begin{table}
\caption{Hurst exponents estimated for the win record time series of $28$
NBA teams in regular seasons 1995-2018 \label{tab:dfa-original}}

\centering{}%
\begin{tabular}{|c|c|c|c|c|c|}
\hline 
team & $H$ & team & $H$ & team & $H$\tabularnewline
\hline 
\hline 
ATL & $0.5223\pm0.0057$ & IND & $0.528\pm0.006$ & PHI & $0.567\pm0.005$\tabularnewline
\hline 
BKN & $0.5485\pm0.0056$ & LAC & $0.5455\pm0.0057$ & PHX & $0.5589\pm0.0066$\tabularnewline
\hline 
BOS & $0.5311\pm0.0067$ & LAL & $0.5365\pm0.0078$ & POR & $0.5936\pm0.0072$\tabularnewline
\hline 
CHI & $0.5456\pm0.0077$ & MEM & $0.5919\pm0.0059$ & SAC & $0.5242\pm0.0073$\tabularnewline
\hline 
CLE & $0.561\pm0.006$ & MIA & $0.5771\pm0.0083$ & SAS & $0.5505\pm0.0078$\tabularnewline
\hline 
DAL & $0.5016\pm0.0051$ & MIL & $0.5265\pm0.0066$ & TOR & $0.559\pm0.005$\tabularnewline
\hline 
DEN & $0.5172\pm0.006$ & MIN & $0.5578\pm0.0066$ & UTA & $0.583\pm0.007$\tabularnewline
\hline 
DET & $0.5186\pm0.0073$ & NYK & $0.5484\pm0.0066$ & WAS & $0.5354\pm0.0065$\tabularnewline
\hline 
GSW & $0.5323\pm0.0066$ & OKC & $0.4960\pm0.0051$ &  & \tabularnewline
\hline 
HOU & $0.5583\pm0.0052$ & ORL & $0.5687\pm0.0068$ &  & \tabularnewline
\hline 
\end{tabular}
\end{table}
Note that for the most of the teams our $H$ estimates are smaller
than those reported in \cite{Ferreira2018PhysA}. This is because
we have fitted $F\left(s\right)$ for $s\in\left[5,70\right]$, which
is narrower interval than the one used in \cite{Ferreira2018PhysA}.
We prefer this narrower interval, because it is still comparatively
broad (spans a bit over one order of magnitude) and the smallest $R^{2}$
over all teams in the data set is one of the largest obtained while
considering other alternative intervals ($R^{2}=0.988$). $R^{2}$
reported in \cite{Ferreira2018PhysA} is smaller which indicates that
our fits are somewhat better. Also from a simple visual comparison
it is pretty clear that the fits obtained in \cite{Ferreira2018PhysA}
are not very good (compare Fig.~\ref{fig:original-extreme} in this
paper and Fig.~1 in \cite{Ferreira2018PhysA}). The slopes reported
in \cite{Ferreira2018PhysA} are somewhat larger than the ones reported
by us, because their fitting interval is broader and includes $s>70$,
while for the most of the teams $F\left(s\right)$ starts curve upwards
for $s>70$, which inflates $H$ estimates. We believe that this curving
upwards might indicate that there is another scaling regime, possibly
related to the long-term decisions made on the season time scale ($82$
games).  This curving upward might either be artifact observed due
to the small sample size, or due to season-to-season persistence.

\section{Checking the empirical results against random models\label{sec:analysis-shuffle}}

For small data sets, such as this one, estimation of the true underlying
Hurst exponent using DFA (or other similar techniques) is not very
reliable. E.g., in \cite{Drozdz2009EPL} MF-DFA, generalization of
DFA, was used to estimate $H$ of synthetic series of various lengths,
it was found that as small as $10^{4}$ points might be sufficient
in some cases. \cite{Drozdz2009EPL} also lists numerous sophisticated
bootstrap techniques, which preserve multifractal properties of the
series. Yet our goals are simpler and for this paper it seems sufficient
to use simple shuffling algorithms and also compare $H$ estimates
against some simple random models. This section is dedicated to these
tests. In all of the following tests our null hypothesis is that the
data is neither persistent nor anti-persistent.

First of all let us shuffle all the games each team has played. This
type of shuffling should remove persistence effects across all time
scales. As can be seen in Fig.~\ref{fig:record-example} the resulting
shuffled win record series (green curve) is nothing alike the original
series (red curve). We have shuffled team records $10^{4}$ times
and have estimated $95\%$ confidence intervals (CIs) for what $H$
values could be reasonably expected to be obtained from a similar,
yet non-persistent, series. If the estimated $H$ values are outside
the obtained CIs, then for those cases we could reject the null hypothesis.

\begin{table}
\caption{Comparing $H$ estimates against CIs obtained from the fully shuffled
data\label{tab:full-shuffle}}

\centering{}%
\begin{tabular}{|c|c|c|c|}
\hline 
Team &  $H$ & Shuffle CI & Reject the null?\tabularnewline
\hline 
\hline 
ATL & $0.5223\pm0.0057$ & $[0.4667,0.5628]$ & No\tabularnewline
\hline 
BKN & $0.5485\pm0.0056$ & $[0.4676,0.5640]$ & No\tabularnewline
\hline 
BOS & $0.5311\pm0.0067$ & $[0.4675,0.5624]$ & No\tabularnewline
\hline 
CHI & $0.5456\pm0.0077$ & $[0.4665,0.5639]$ & No\tabularnewline
\hline 
CLE & $0.561\pm0.006$ & $[0.4657,0.5625]$ & No\tabularnewline
\hline 
DAL & $0.5016\pm0.0051$ & $[0.4670,0.5630]$ & No\tabularnewline
\hline 
DEN & $0.5172\pm0.006$ & $[0.4663,0.5637]$ & No\tabularnewline
\hline 
DET & $0.5186\pm0.0073$ & $[0.4666,0.5637]$ & No\tabularnewline
\hline 
GSW & $0.5323\pm0.0066$ & $[0.4673,0.5635]$ & No\tabularnewline
\hline 
HOU & $0.5583\pm0.0052$ & $[0.4669,0.5626]$ & No\tabularnewline
\hline 
IND & $0.528\pm0.006$ & $[0.4658,0.5640]$ & No\tabularnewline
\hline 
LAC & $0.5455\pm0.0057$ & $[0.4661,0.5632]$ & No\tabularnewline
\hline 
LAL & $0.5365\pm0.0078$ & $[0.4664,0.5640]$ & No\tabularnewline
\hline 
MEM & $0.5919\pm0.0059$ & $[0.4675,0.5631]$ & \textbf{Yes}\tabularnewline
\hline 
MIA & $0.5771\pm0.0083$ & $[0.4671,0.5631]$ & \textbf{Yes}\tabularnewline
\hline 
MIL & $0.5265\pm0.0066$ & $[0.4660,0.5622]$ & No\tabularnewline
\hline 
MIN & $0.5578\pm0.0066$ & $[0.4667,0.5633]$ & No\tabularnewline
\hline 
NYK & $0.5484\pm0.0066$ & $[0.4672,0.5639]$ & No\tabularnewline
\hline 
OKC & $0.4960\pm0.0051$ & $[0.4674,0.5630]$ & No\tabularnewline
\hline 
ORL & $0.5687\pm0.0068$ & $[0.4674,0.5640]$ & No\tabularnewline
\hline 
PHI & $0.567\pm0.005$ & $[0.4668,0.5635]$ & No\tabularnewline
\hline 
PHX & $0.5589\pm0.0066$ & $[0.4658,0.5628]$ & No\tabularnewline
\hline 
POR & $0.5936\pm0.0072$ & $[0.4674,0.5626]$ & \textbf{Yes}\tabularnewline
\hline 
SAC & $0.5242\pm0.0073$ & $[0.4661,0.5639]$ & No\tabularnewline
\hline 
SAS & $0.5505\pm0.0078$ & $[0.4684,0.5641]$ & No\tabularnewline
\hline 
TOR & $0.559\pm0.005$ & $[0.4667,0.5627]$ & No\tabularnewline
\hline 
UTA & $0.583\pm0.007$ & $[0.4676,0.5636]$ & \textbf{Yes}\tabularnewline
\hline 
WAS & $0.5354\pm0.0065$ & $[0.4666,0.5631]$ & No\tabularnewline
\hline 
\end{tabular}
\end{table}
As we can see in Table~\ref{tab:full-shuffle} we are able reject
the null hypothesis in $4$ cases: for MEM (Vancouver/Memphis Grizzlies),
MIA (Miami Heat), POR (Portland Trail Blazers) and UTA (Utah Jazz)
the estimated $H$ is larger than would be expected if the series
would be non-persistent. When analyzing $28$ cases under $5\%$ significance
level we could reasonably expect to have $1$ or $2$ false rejections.
Though these expectations do not account for the fact that the empirical
series under consideration are not mutually independent, as a win
for one team is a loss for another. In this context obtaining $4$
false rejections no longer seems unexpected. Thus we consider the
obtained evidence questionable.

We can also check whether the streak lengths (first passage times)
of the original and the shuffled series come from the different distributions.
Note that for this test we use not the win record series (shown in
Fig.~\ref{fig:record-example}), but the binary won-lost series.
For the sake of consistency for this test we use only one shuffled
time series per team. Here we use Kolmogorov-Smirnov test \cite{Lopes2011Springer}
with significance level $\alpha=0.05$. For the both winning and losing
streaks we were able reject the null hypothesis (that the sets of
streaks come from the same distribution) only for the winning streaks
of MEM. As these results is within expected false rejection count,
we consider the obtained evidence is negligible.

We can perform another test using the streak length distribution:
whether the streak lengths from the original data are inconsistent
with an independent randomness model. For this test we assume the
following model: the team has probability $p_{w}$ to win each game
independently of the other games and $p_{w}$ is estimated based on
the overall win-loss record of the team. Once again here we have used
only one randomly generated series. This time using Kolmogorov-Smirnov
test ($\alpha=0.05$) we were able to reject the null hypothesis only
for the winning streaks of LAL (Los Angeles Lakers). We consider this
evidence to be negligible, as the rejection count is within the expected
false rejection count.

This independent randomness model could be further simplified by ignoring
the overall win-loss record (setting $p_{w}=0.5$ for all teams).
But in this case the number of rejections is significantly higher:
$4$ rejections for the winning streaks and $5$ rejections for the
losing streaks. The increased rejection count is expected for the
teams which won noticeably more or less than half of its games. What
is striking is that the fully random model is consistent with records
of the most teams, which could indicate that the technique is not
sensitive enough to capture small differences in the probability to
win a single game.

To further our discussion let us try a slightly different shuffling
algorithm. This time let us shuffle the games inside the seasons,
but preserve the ordering of the seasons. The in-season algorithm
destroys the possible game-to-game persistence, but retains season-to-season
persistence. An example of the series shuffled using this algorithm
(blue curve) is shown in Fig.~\ref{fig:record-example}. As can be
easily seen the blue curve repeats the general shape of the original
(red) curve. Once again we have obtained $10^{4}$ shuffles and have
estimated $95\%$ CIs for $H$ estimates. As we can see in Table~\ref{tab:other-shuffle}
we are able to reject the null hypothesis in just two cases (POR and
UTA). As these results is within expected false rejection count, we
consider the obtained evidence is negligible.

\begin{figure}
\begin{centering}
\includegraphics[width=0.7\textwidth]{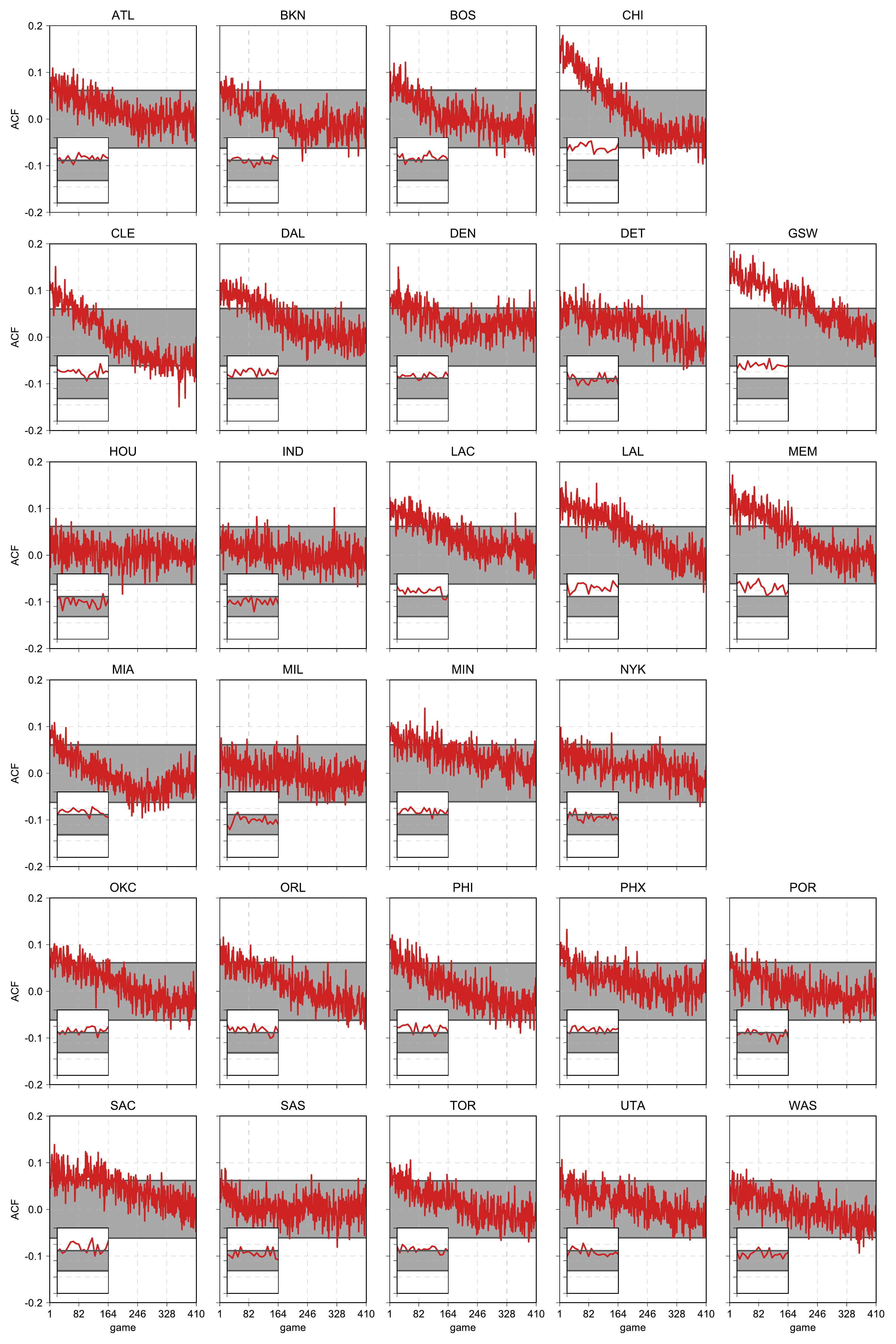}
\par\end{centering}
\caption{(color online) Comparison between the autocorrelation functions of
the empirical series (red curves) and the $95\%$ confidence intervals
obtained by fully shuffling the respective empirical series (gray
areas). Embedded subplots provide a zoom in on the x-axis (to $[1,20]$
games interval).\label{fig:acf-total}}
\end{figure}

\begin{figure}
\begin{centering}
\includegraphics[width=0.7\textwidth]{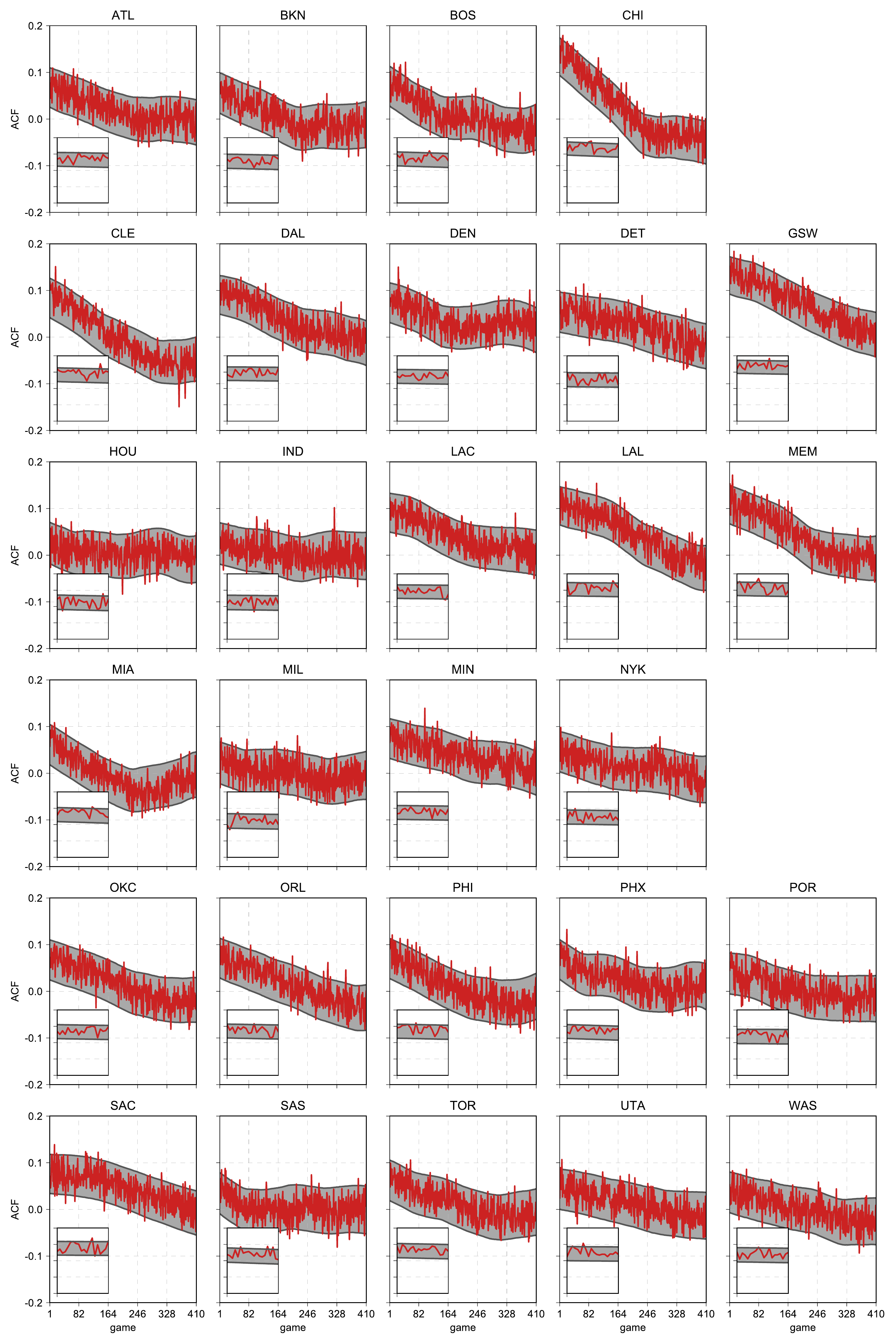}
\par\end{centering}
\caption{(color online) Comparison between the autocorrelation functions of
the empirical series (red curves) and the $95\%$ confidence intervals
obtained by shuffling the respective empirical series in-season (gray
areas). Embedded subplots provide a zoom in on the x-axis (to $[1,20]$
games interval).\label{fig:acf-inseason}}
\end{figure}

\begin{table}
\caption{Comparing $H$ estimates against CIs obtained from the in-season shuffled
data\label{tab:other-shuffle}}

\centering{}%
\begin{tabular}{|c|c|c|c|}
\hline 
team & $H$ & In-season shuffle CI & Reject the null?\tabularnewline
\hline 
\hline 
ATL & $0.5223\pm0.0057$ & $[0.4925,0.5825]$ & No\tabularnewline
\hline 
BKN & $0.5485\pm0.0056$ & $[0.5067,0.5947]$ & No\tabularnewline
\hline 
BOS & $0.5311\pm0.0067$ & $[0.5187,0.6043]$ & No\tabularnewline
\hline 
CHI & $0.5456\pm0.0077$ & $[0.5284,0.6129]$ & No\tabularnewline
\hline 
CLE & $0.561\pm0.006$ & $[0.5194,0.6056]$ & No\tabularnewline
\hline 
DAL & $0.5016\pm0.0051$ & $[0.4773,0.5707]$ & No\tabularnewline
\hline 
DEN & $0.5172\pm0.006$ & $[0.4916,0.5839]$ & No\tabularnewline
\hline 
DET & $0.5186\pm0.0073$ & $[0.4834,0.5736]$ & No\tabularnewline
\hline 
GSW & $0.5323\pm0.0066$ & $[0.4975,0.5859]$ & No\tabularnewline
\hline 
HOU & $0.5583\pm0.0052$ & $[0.4968,0.5853]$ & No\tabularnewline
\hline 
IND & $0.528\pm0.006$ & $[0.4853,0.5776]$ & No\tabularnewline
\hline 
LAC & $0.5455\pm0.0057$ & $[0.4885,0.5835]$ & No\tabularnewline
\hline 
LAL & $0.5365\pm0.0078$ & $[0.4910,0.5820]$ & No\tabularnewline
\hline 
MEM & $0.5919\pm0.0059$ & $[0.4963,0.5867]$ & No\tabularnewline
\hline 
MIA & $0.5771\pm0.0083$ & $[0.5125,0.5995]$ & No\tabularnewline
\hline 
MIL & $0.5265\pm0.0066$ & $[0.4892,0.5793]$ & No\tabularnewline
\hline 
MIN & $0.5578\pm0.0066$ & $[0.4881,0.5795]$ & No\tabularnewline
\hline 
NYK & $0.5484\pm0.0066$ & $[0.4886,0.5800]$ & No\tabularnewline
\hline 
OKC & $0.4960\pm0.0051$ & $[0.4917,0.5810]$ & No\tabularnewline
\hline 
ORL & $0.5687\pm0.0068$ & $[0.4951,0.5858]$ & No\tabularnewline
\hline 
PHI & $0.567\pm0.005$ & $[0.4995,0.5895]$ & No\tabularnewline
\hline 
PHX & $0.5589\pm0.0066$ & $[0.5147,0.5994]$ & No\tabularnewline
\hline 
POR & $0.5936\pm0.0072$ & $[0.4828,0.5768]$ & \textbf{Yes}\tabularnewline
\hline 
SAC & $0.5242\pm0.0073$ & $[0.4752,0.5688]$ & No\tabularnewline
\hline 
SAS & $0.5505\pm0.0078$ & $[0.5028,0.5899]$ & No\tabularnewline
\hline 
TOR & $0.559\pm0.005$ & $[0.5000,0.5898]$ & No\tabularnewline
\hline 
UTA & $0.583\pm0.007$ & $[0.4814,0.5744]$ & \textbf{Yes}\tabularnewline
\hline 
WAS & $0.5354\pm0.0065$ & $[0.4893,0.5781]$ & No\tabularnewline
\hline 
\end{tabular}
\end{table}
Another way to look at temporal persistence is to examine ACF of the
series. In Fig.~\ref{fig:acf-total} we see that most of the teams
have weakly positive ACFs for the lags up to $82$ games (one season).
In quite a few cases, most notably CHI (Chicago Bulls) and GSW (Golden
State Warrriors), the ACF is outside reasonably expected range (gray
area) if the series would be random (comparison is made against the
fully shuffled series). Yet the excess autocorrelation is well explained
by taking season-to-season variations into account. In Fig.~\ref{fig:acf-inseason}
the comparison is made against the in-season shuffled data and as
one can see the empirical ACFs are now within the reasonably expected
range (gray area).


\section{Conclusion\label{sec:Conclusion}}

In our exploration of the NBA regular season data set from the period
1995-2018 we have not found enough evidence supporting the ``winning
streak'' phenomenon. Although for the most of the teams we have obtained
Hurst exponents larger than $0.5$ (these results are somewhat consistent
with the ones reported in \cite{Ferreira2018PhysA}), which would
indicate the presence of persistence, these results are mostly consistent
with the considered random models (full data shuffle, in-season data
shuffle and independent randomness model) at significance level of
$5\%$. It appears that the observed persistence is simply an illusion
caused by the limited amount of data under consideration. Nevertheless
DFA, as suggested by \cite{Ferreira2018PhysA}, seems to be an interesting
tool to be applied to the sports time series and extensive application
of it on a long time series could potentially provide arguments to
the ongoing ``hot hand'' debate. Furthermore it seems that DFA could
be also used to answer different question related to the different
manegerial strategies teams use, e.g., in \cite{Rotundo2015PhysA}
it was examined whether religious practicies have impact on the baby
birth times.

We have backed the results obtained using DFA technique by analyzing
streak length distributions. Namely, using Kolmogorov-Smirnov test
($\alpha=0.05$), we were unable to find any evidence indicating that
the empirical streak lengths would be inconsistent with the independent
randomness model. These results also suggest that there is no evidence
for the persistence in winning record, or the ``winning streak'',
phenomenon.

Analysis carried out using ACFs paints a more sophisticated picture.
Empirical ACFs are not fully consistent with ACFs one would expect
if there were no persistence. Yet these inconsistencies are well explained
by retaining season-to-season variations. This would point to an obvious
fact that long-term managerial decisions have an impact on team's
performance. While on the other hand short-term decisions and effects
as ``wining streaks'' do not seem to have effects which would be inconsistent
with random models.

Finally we would like to advise against gambling by relying on the
persistence. Not only because we did not find sufficient evidence
for the game-to-game persistence phenomenon in the NBA time series,
but also because in the cases where persistence was detected (e.g.,
financial time series), there are still no reliable forecasting algorithms
relying on the persistence alone. Gambling on the other hand could
also influence the outcome of the games, but we would doubt if it
had effect in as reputable league as NBA. It is likely that DFA or
simple random models (similar to ones considered here) could be used
to detect betting frauds in less reputable sports leagues.

We have made all of the scripts as well as the data, we have used
in this analysis, freely available using GitHub \cite{Kononovicius2018GitNBA}.


\end{document}